\documentclass[square]{ws-procs11x85}
\usepackage{balance} 

\def\Journal#1#2#3#4{{#1} {\bf #2}, #3 (#4)}

\newcommand{\met}{\hbox{E\kern-0.5em\lower-0.1ex\hbox{/}}_T}

\begin{document}

\twocolumn[
\title{Formation of relativistic MHD jets: 
       stationary state solutions \& numerical simulations}

\author{Christian Fendt}
\address{Max Planck Institute for Astronomy, K\"onigstuhl 17, D-69117 Heidelberg, Germany\\E-mail: fendt@mpia.de}

\author{Elisabetta Memola}
\address{INAF - Osservatorio Astronomico di Brera, via Brera 28, 20121 Milano, Italy}


\begin{abstract}
We discuss numerical results of relativistic magnetohydrodynamic (MHD) jet formation models.
We first reviews some examples of  stationary state solutions treating the collimation and
acceleration process of relativistic MHD jets.
We provide an {\it a posteriori} check for the MHD condition in highly magnetized flows,
namely the comparison of particle density to Goldreich-Julian density.
Using the jet dynamical parameters calculated from the MHD model we show the
rest-frame thermal X-ray spectra of the jet, from which we derive the overall
spectrum taking into account a variation of Doppler boosting and Doppler shift of emission 
lines along the outflow.
Finally, we present preliminary results of relativistic MHD simulations of jet
formation demonstrating the acceleration of a low velocity (0.01\,c) disk wind
to a collimated high velocity (0.8\,c).
\end{abstract}
%
 \vskip28pt 
]

\bodymatter

\section{Introduction}
Astrophysical jets are commonly believed to consist mainly of collimated disk winds,
launched, accelerated, collimated by magnetic forces
(see \cite{blan82, pudr83, came86, besk97, ouye97, heyv03, pudr07, komi07}).
The jet is initiated as a slow wind by a process which is not yet completely understood,
in particular its time-dependent character.
Most probably, some disk instability is responsible for adjusting the mass flux from
disk into an outflow in the direction perpendicular to the disk surface.
The disk wind is first launched magneto-centrifugally and further accelerated and
(self-) collimated into a narrow beam by Lorentz forces.
It should be noted, however, that certain important properties of relativistic jets 
are still purely known.
We don't know whether relativistic jets are launched as leptonic or hadronic beams.
We do not know much about the intrinsic jet velocities.
We also don't know the magnetic field structure of the jet or of its source.

Traditionally, the theoretical task to settle these issues can be divided into five 
fundamental questions:
\begin{itemlist}
\item
The question of how to collimate and accelerate a disk/stellar wind into a jet (jet formation)?
\item
The problem of how to eject the disk/stellar material into the wind (jet launching)?
\item
The question of the accretion disk structure and evolution? 
\item
The question of the origin of the governing magnetic field?
\item
The question of the jet propagation, its interaction with 
   ambient medium, its stability and the radiative signatures involved.
\end{itemlist}
In the end, it would be desirable to obtain fully self-consistent 
solutions of relativistic MHD jets - however, such solutions have not 
yet been obtained in the stationary case.
However, the progress in numerical methods and computer power now allows
for time-dependent simulations of relativistic jet formation.

In this work we concentrate on the first problem, the structure and dynamics of the
jet formation region close to the source.
We present stationary state and time-dependent numerical results of relativistic jets
as collimating disk winds.
For alternative or complementary viewpoints 
(e.g. electromagnetic jet formation, Blandford-Znajek mechanism) we refer to the literature.

\section{Concepts of relativistic MHD jets}
Considerations of stationary MHD flows have revealed that relativistic jets
must be strongly magnetized \cite{mich69, came86, li93, fend96}.
In that case, the available magnetic energy can be transfered into a small
amount of mass with high kinetic energy.
On the other hand, a very strong magnetization may be in conflict with the
MHD assumption lacking a sufficient large amount of electric charges
which are needed to drive the electric current system.

Theoretical modeling of jet formation requires to solve the governing
MHD equations.
However, due to the complexity of MHD and the astrophysical boundary conditions
indicated, a completely self-consistent MHD solution for the jet formation
process being compatible with all the generic features
(MHD self-collimation, accretion-ejection mechanism,
magnetic field generation, spatial and time scales  etc.)
does not yet exist.

We believe that the jet formation region may fairly well be approximated as 
{\em axisymmetric}. 
Non-axisymmetric distortions may actually hinder the formation of 
powerful jets as probably demonstrated by the existence of many strongly magnetized,
rapidly rotating accretion disk systems which do not have jets
(e.g. cataclysmic variables or most pulsars).

Compared to the non-relativistic case, in relativistic MHD some specific features 
arise affecting both the model approach and the physical mechanism of jet formation.
We first mention the {\em light cylinder} of the magnetosphere.
If $\Omega_{\rm F}$ is interpreted as angular velocity of magnetic field
lines, at the light cylinder radius $R_{\rm L} = c/\Omega_{\rm F}$ of that 
field line its rotational velocity equals the speed of light. 
In general relativity, frame dragging adds another, inner, light "cylinder" 
to the system of MHD equations.
For vanishing mass density, the location of the MHD Alfv\'en surface
approaches the light surface.
Another relativistic MHD feature is that the {\em electric field} cannot be 
neglected.
The poloidal electric field component $\vec{E}_{\perp}$ (which is 
perpendicular to the poloidal magnetic field)
reaches considerable strength, $E_{\perp} = (R/R_{\rm L}) B_{\rm p}$.
The electric field pressure gradient and tension adds to the jet collimation 
or de-collimation.

We like to note that neither the light surface nor the relativistic Alfv\'en
surface obeys a self-similar structure (see e.g. \cite{li93}).
Thus, steady-state self-similar relativistic MHD models while having the 
advantage of treating the matter-field interrelation self-consistently 
(compared e.g. to force-free models discussed below) may not necessarily 
take into account all relativistic properties of the flow.

\begin{figure}[t]
\center
\centerline{\psfig{figure=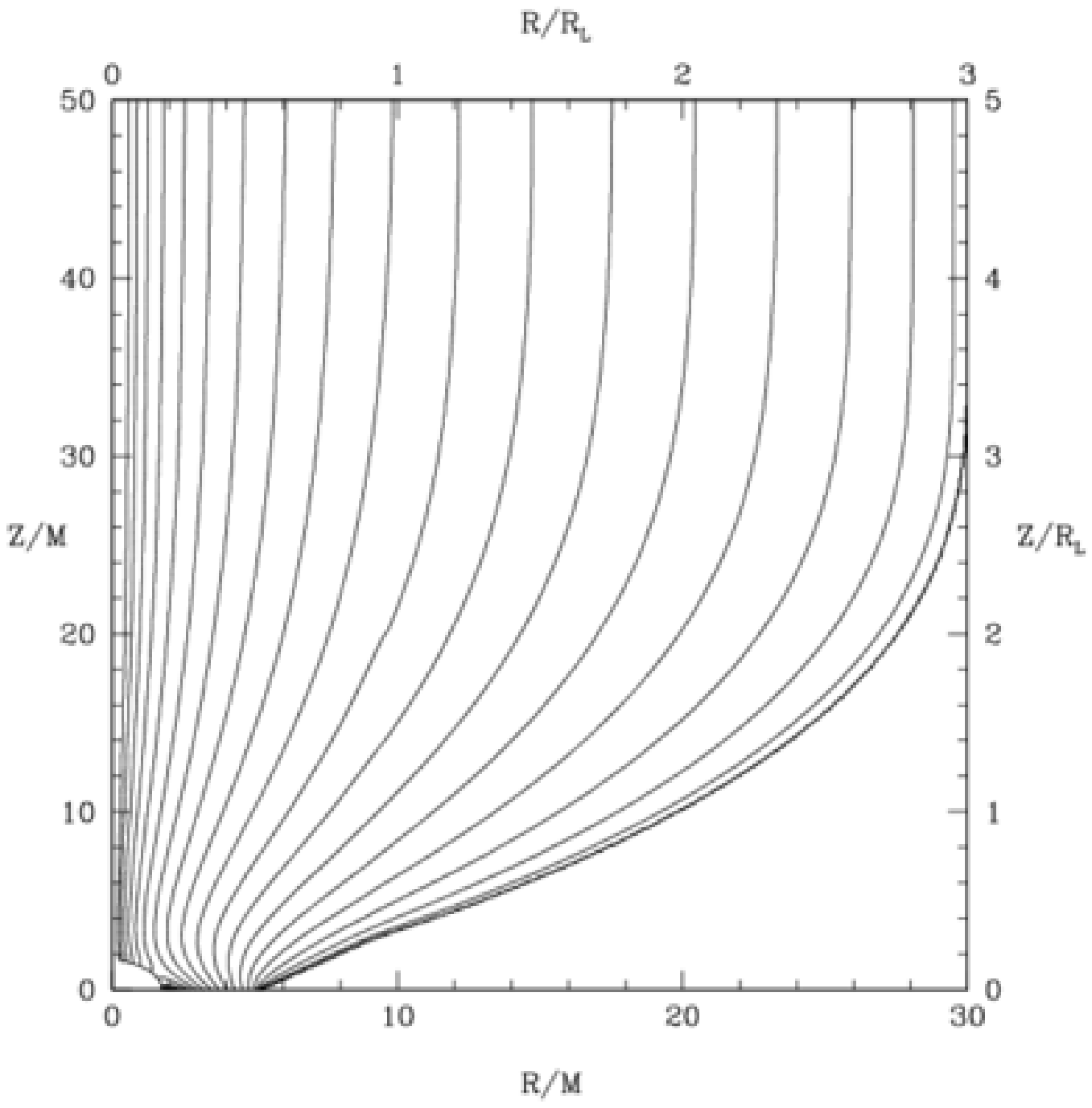,width=8truecm}}

\centerline{\psfig{figure=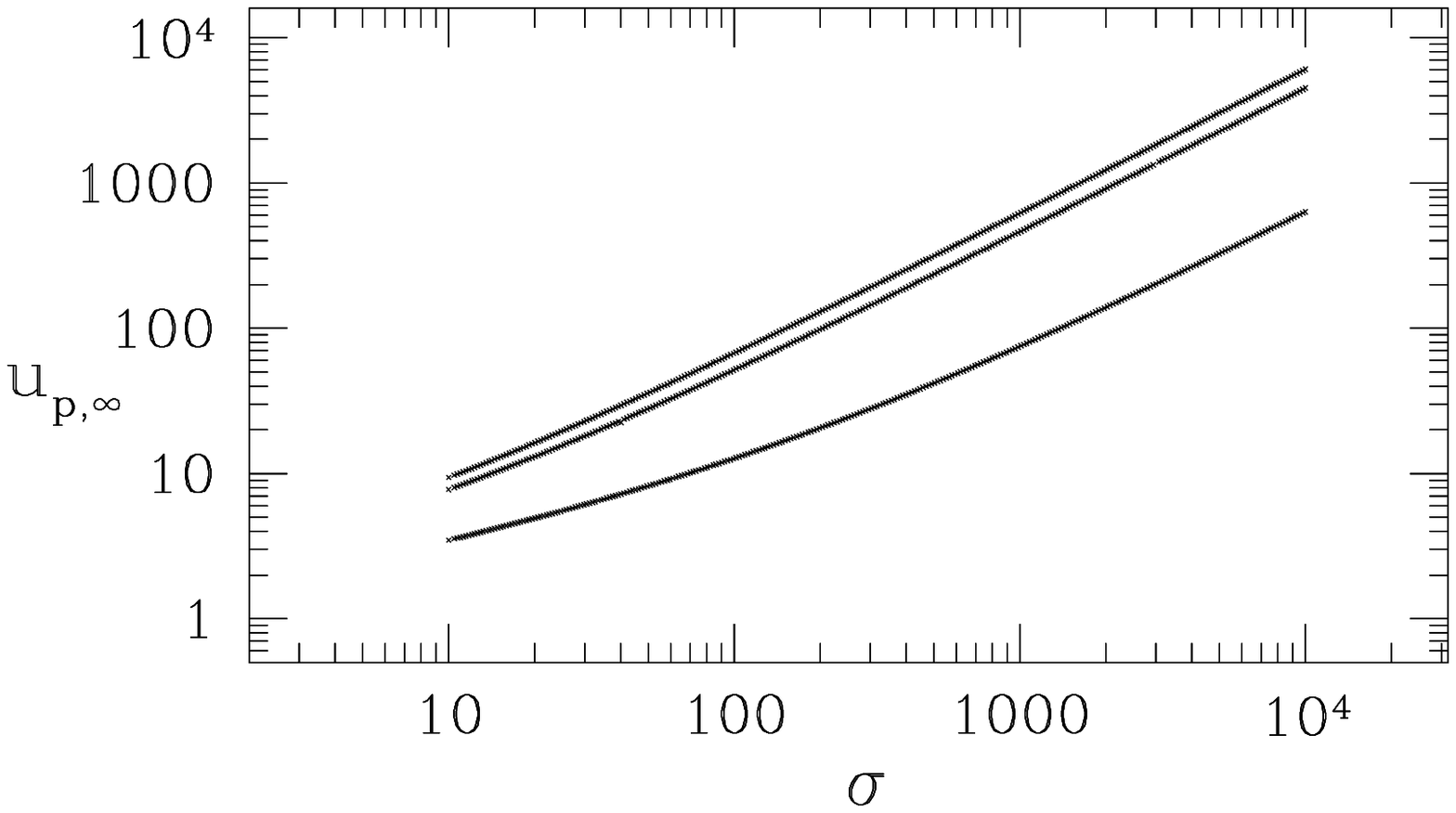,width=8truecm}}

\caption{Solutions of the stationary state axisymmetric MHD equations in
Kerr metric.
Axisymmetric poloidal magnetic field structure ({\em top}, see \cite{fend97}),
cylindrical coordinates normalized to gravitational radii. The inner
boundary coincides with the inner light surface.
Maximum poloidal velocity along a collimating flux surface $u_{\rm p, \infty}$ 
as a function
of flow magnetization $\sigma$ ({\em bottom}, see \cite{fend01a}), 
both in Kerr metric.
\label{fig:jet_stat1}}
\end{figure}

\begin{figure}[t]
\center
\centerline{\psfig{figure=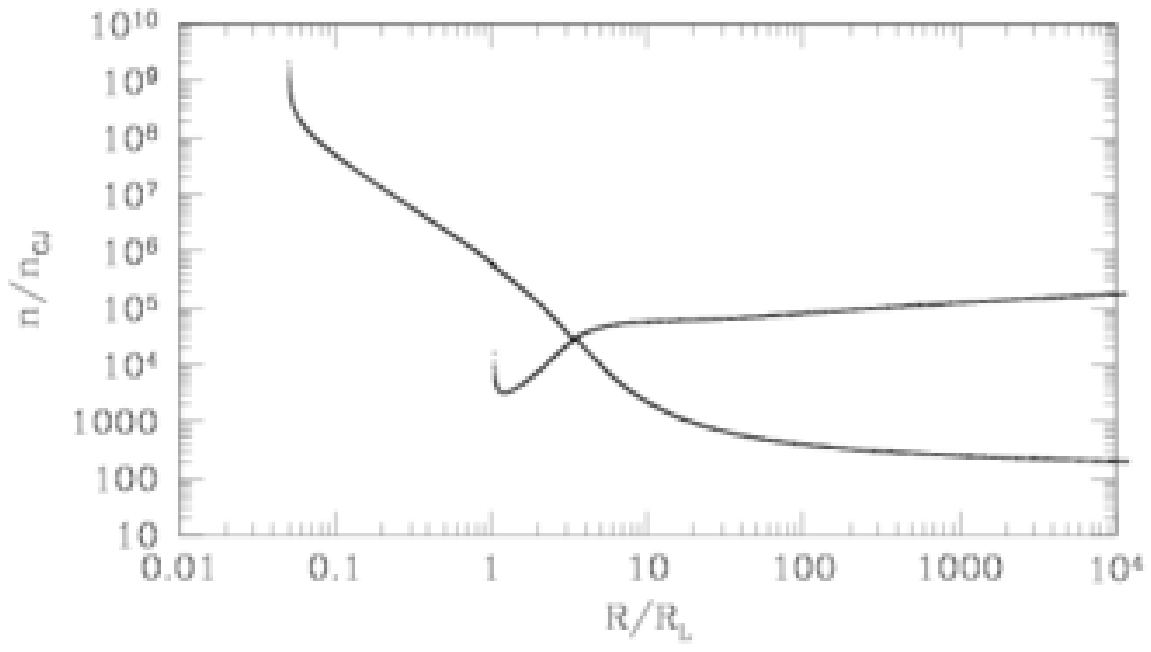,width=8truecm}}
\centerline{\psfig{figure=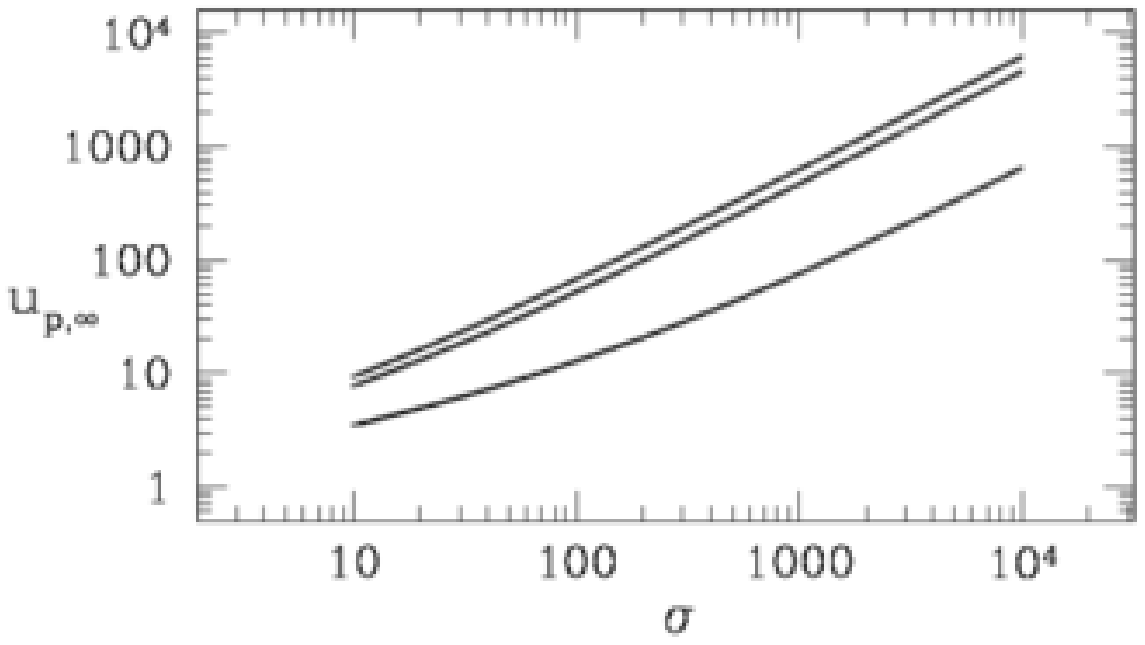,width=8truecm}}
\caption{Solutions of the stationary state relativistic MHD wind equation
for a collimation field structure.
Shown is the ratio of particle density $n$ 
to Goldreich-Julian space charge density  $n_{\rm GJ}$ versus cylindrical
radius along the flux surface (see \cite{fend04})
for examples of highly magnetized flows ({\em top}) with
$\sigma = 1000$ and $\sigma = 5000$ ({\em bottom}).
The physical branch of the wind solution is the one which starts with
high value at small radius and decreases with radius.
\label{fig:jet_stat2}}
\end{figure}

\section{Stationary relativistic MHD jets}\label{sec:relMHD}
In stationary limit {\em global} solutions for the relativistic jet 
structure can be derived (see e.g. \cite{li93, cont94, fend97, meli06}) 
on spatial scales and with a resolution which cannot (yet) be 
reached by time-dependent simulations \cite{kudo98, koid98, mcki07}.
Time-dependent MHD codes usually have also difficulties with low plasma-$\beta$, 
and/or high magnetization - exactly the conditions which generate highly 
relativistic jets.
On the other hand, jets are non-steady phenomena as clearly seen by their 
knotty pattern.
 
Example solutions of stationary state (general) relativistic MHD are
shown in Fig.~\ref{fig:jet_stat1}. 
The figure shows the global axisymmetric field structure of the jet
collimation region considering the (force-free) local cross-field
force-balance (see \cite{fend97}). 
This is a truly 2.5-D solution, taking into account special relativistic
and general relativistic features 
(light surfaces, gravitational redshift, frame dragging).
The asymptotic structure (for large $z$) matches perfectly with the 
analytical 1-D solution. 
The electric current is conserved along the field lines.
The global field solution (inside and outside the light cylinder) is
determined by matching by the regularity condition (along the light cylinder).
Essentially, the field structure is collimating rapidly.
Similar solutions were obtained for Minkowski case, however assuming
differentially rotating foot points of field lines \cite{fend02a}. 
In that case, the light surface has to be determined iteratively as its 
location and shape is not known {\it a priori\/} since both depend on 
the magnetic flux distribution.

Figure~\ref{fig:jet_stat1} also shows an example for the flow dynamics 
along collimating field lines, 
considering the force-balance along the flow and taking into
account magnetic and inertial forces and also gas pressure.
Shown is the maximum poloidal velocity along a field line
(taken at a certain radius) as a function of the flow magnetization
(see \cite{fend04}). 
Essentially, the velocity-magnetization relation follows a variation of
the well-known Michel scaling \cite{mich69}.
This example solution was calculated for hypothetical jets in GRBs.
For high magnetization, bulk Lorentz factors of 1000 were obtained.
The transition between almost conical field lines (split-monopole) 
close to the source and asymptotic cylindrical field lines leads to
an interesting behavior.
As these flux surfaces actually re-collimate (see the separation between
the flux surfaces), for high magnetization we do not find stationary
solutions for each radius beyond the re-collimation distance. 
This indicates that the flow is highly time-variable, giving rise to
internally differential motions and thus internal shocks which may
trigger the prompt emission of the GRB (see \cite{fend04}).

\begin{figure}[t]
\center
\centerline{\psfig{figure=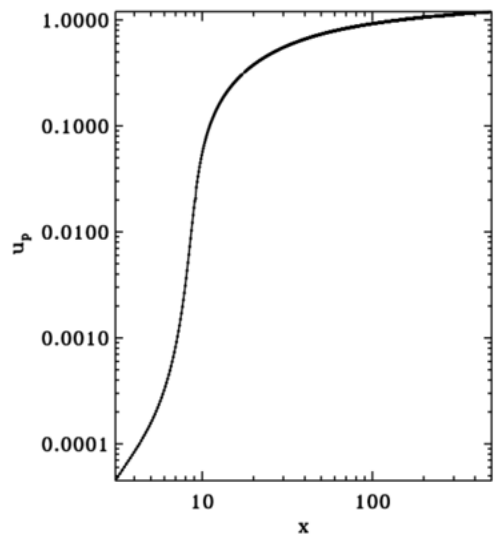,width=7truecm}}

\centerline{\psfig{figure=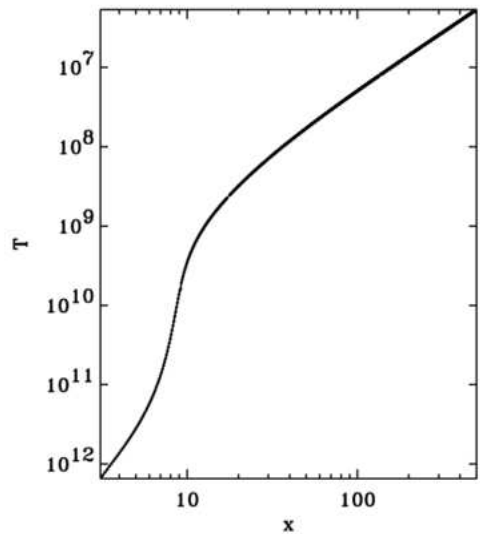,width=7truecm}}

\caption{Thermal X-ray spectra from relativistic MHD jets.
Shown are signatures of the underlying magnetohydrodynamic
structure - the poloidal jet velocity $u_{\rm p} \equiv \gamma v_{\rm p}$ 
along the flux surface
({\em top}) and the gas temperature $T$ ({\em bottom}),
see \cite{fend01a,memo02}. Distance $x$ is normalized to
the gravitational radius.
\label{fig:xray1}}
\end{figure}

\begin{figure}[t]
\center
\centerline{\psfig{figure=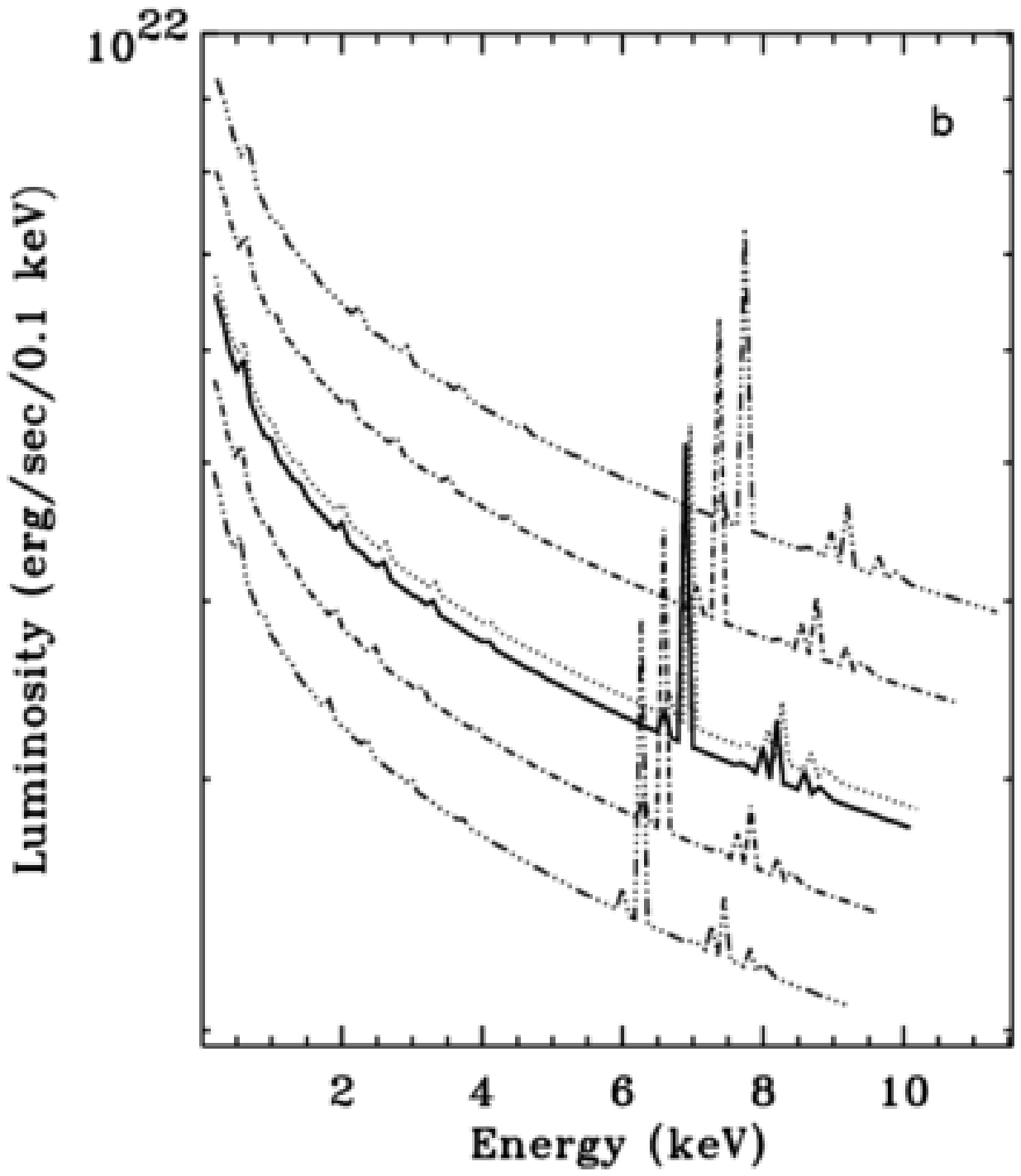,width=7truecm}}

\centerline{\psfig{figure=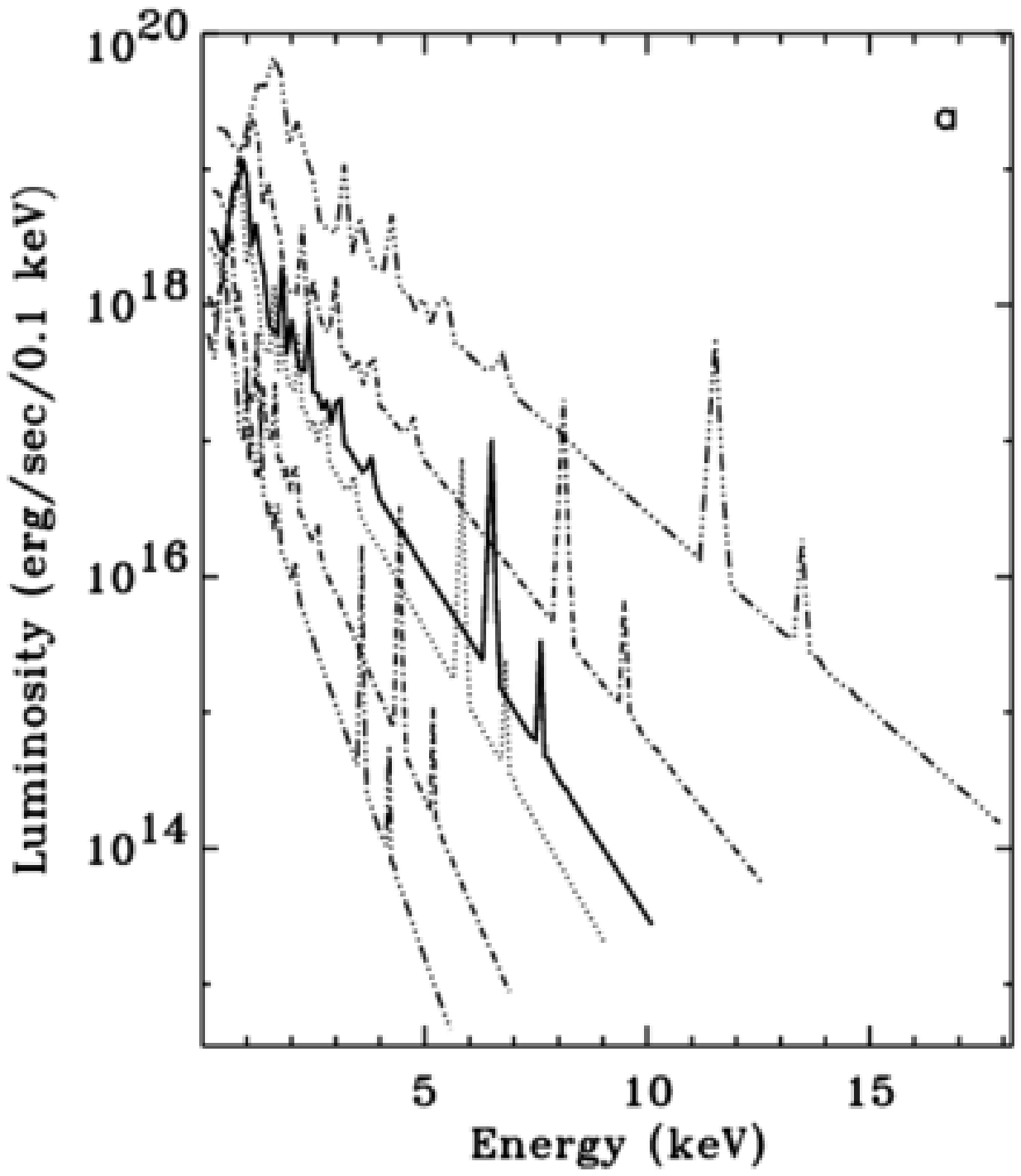,width=7truecm}}
\caption{Thermal X-ray spectra from relativistic MHD jets.
Shown are example spectra derived for the accelerating
hot gas. The {\em top} panel shows a hot ($10^9$\,K), slow gas
component, the {\em lower} panel a cooler ($10^7$\,K), but
faster gas component\cite{memo02}. The different lines show
the spectra for different viewing angles ($-40, -20, 0, +20, +40$
degrees).
\label{fig:xray2}}
\end{figure}

\section{Relativistic jets \& MHD assumption}\label{sec:assumeMHD}
An essential issue for relativistic MHD models is the question whether
the MHD limit actually holds for the parameter space investigated.
A small mass load implies a vanishing amount of charge carries.
Electric currents may not longer be established.
The validity of the MHD condition can be estimated by comparing the 
particle density $n$ in respect to the Goldreich-Julian density 
$n_{\rm GJ}$ 
(see \cite{usov94,fend04}; see \cite{mela96} for another approach).
Figure~\ref{fig:jet_stat2} demonstrates our {\it a posteriori\/} check for 
the MHD condition for highly magnetized flows.
For large radii the fraction $({n}/{n_{\rm GJ}})$ becomes constant value 
$\sim 100 >> 1$, telling us that our MHD solution for this choice 
of parameters is consistent with the MHD assumption. 
The solutions shown in Fig.~\ref{fig:jet_stat2} were derived for collimating 
field lines with slightly diverging flux, 
$B_{\rm p} r^2 \equiv \Phi(r) \sim r^q$ with $q \sim 0.1 > 0$. 
For a more diverging field structure (lower $q$),
a radius will exist beyond which the MHD condition will break down,
$({n}/{n_{\rm GJ}}) \simeq 1$.

\begin{figure*}[t]
\center
\centerline{\psfig{figure=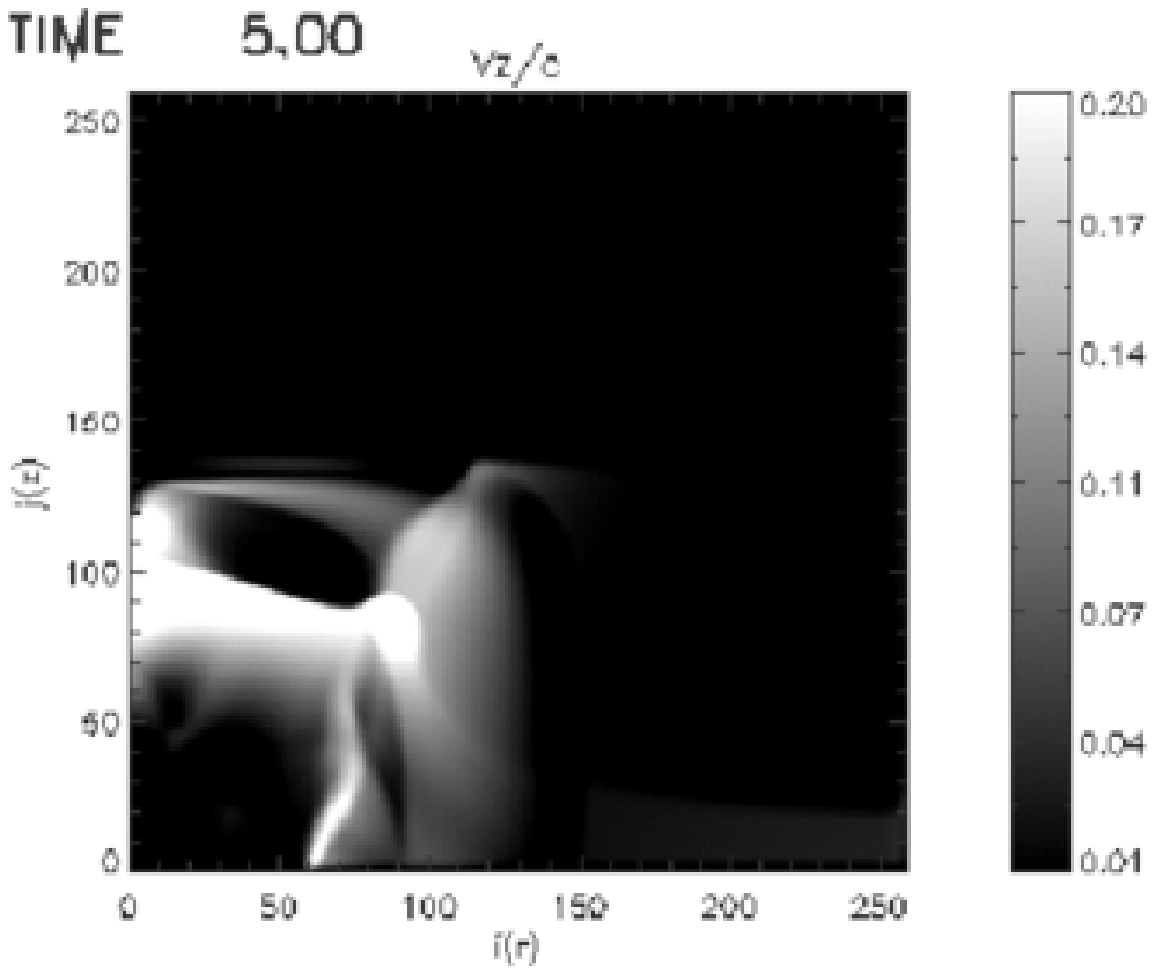,width=9truecm}
            \psfig{figure=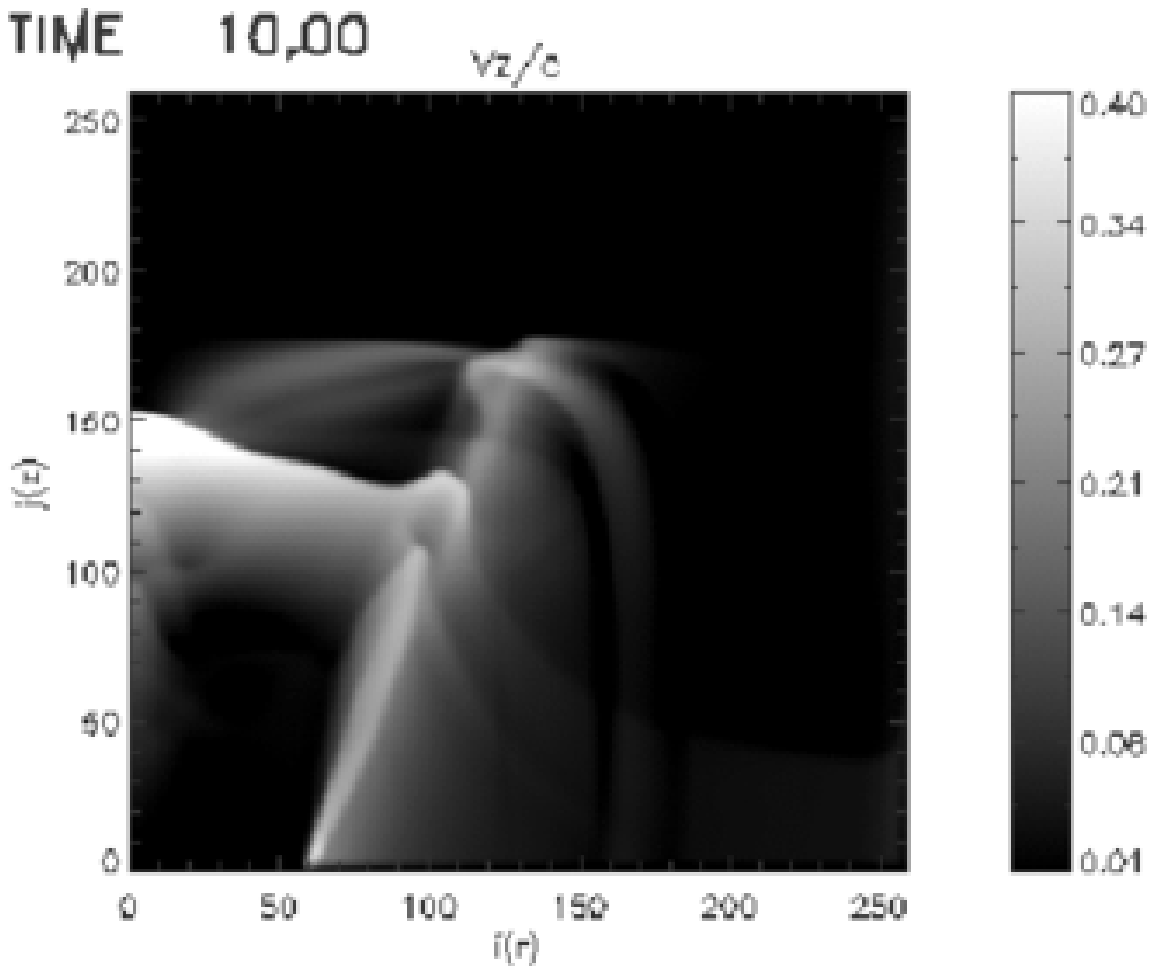,width=9truecm}}
\centerline{\psfig{figure=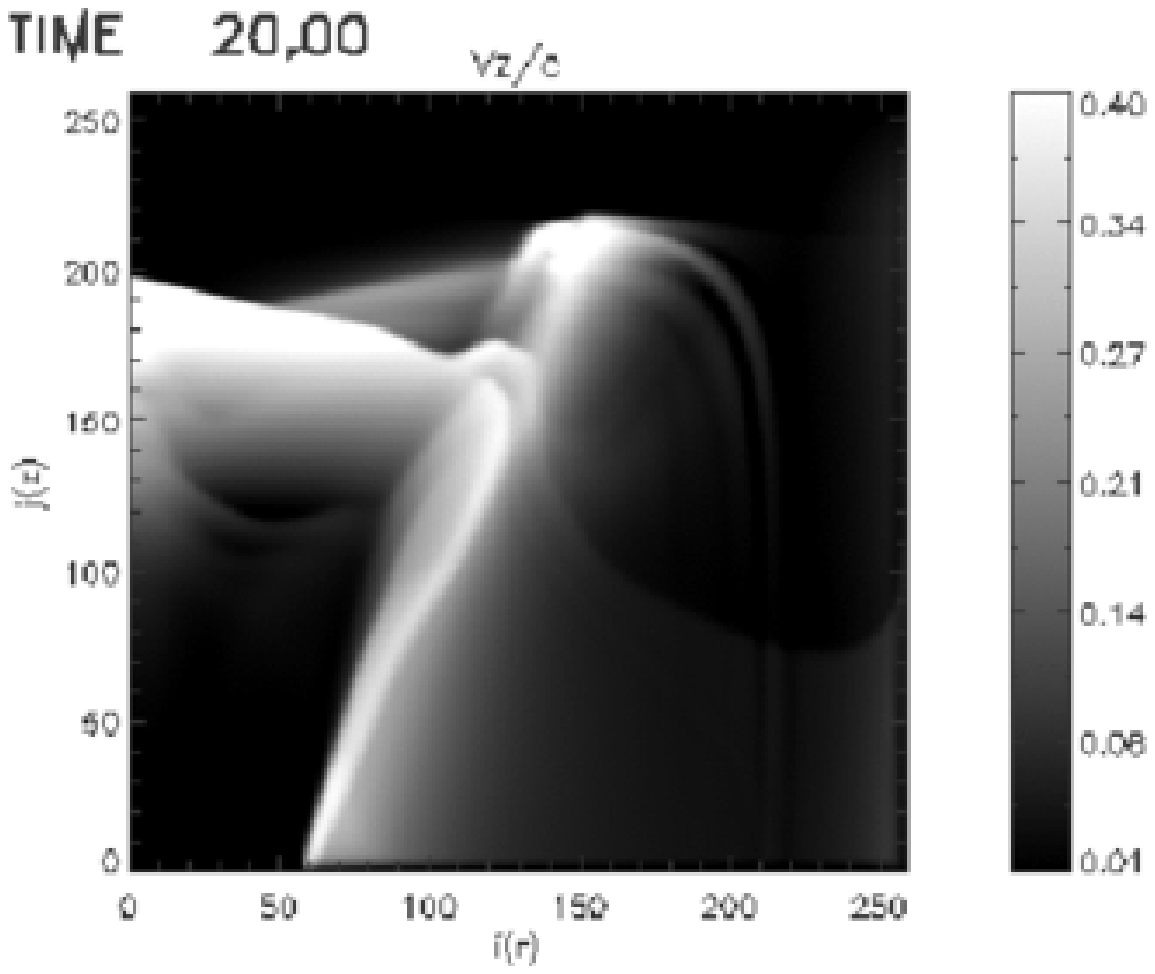,width=9truecm}
            \psfig{figure=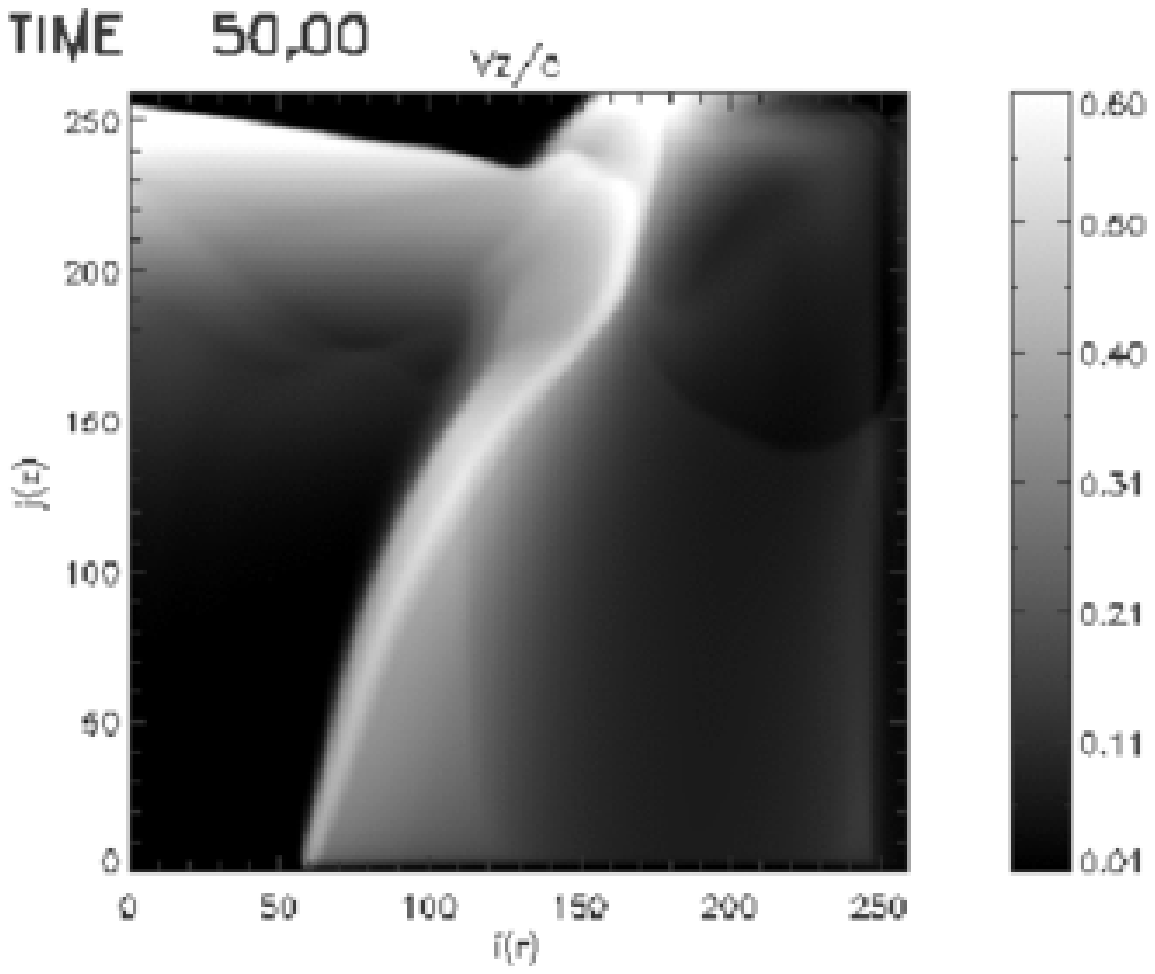,width=9truecm}}
\centerline{\psfig{figure=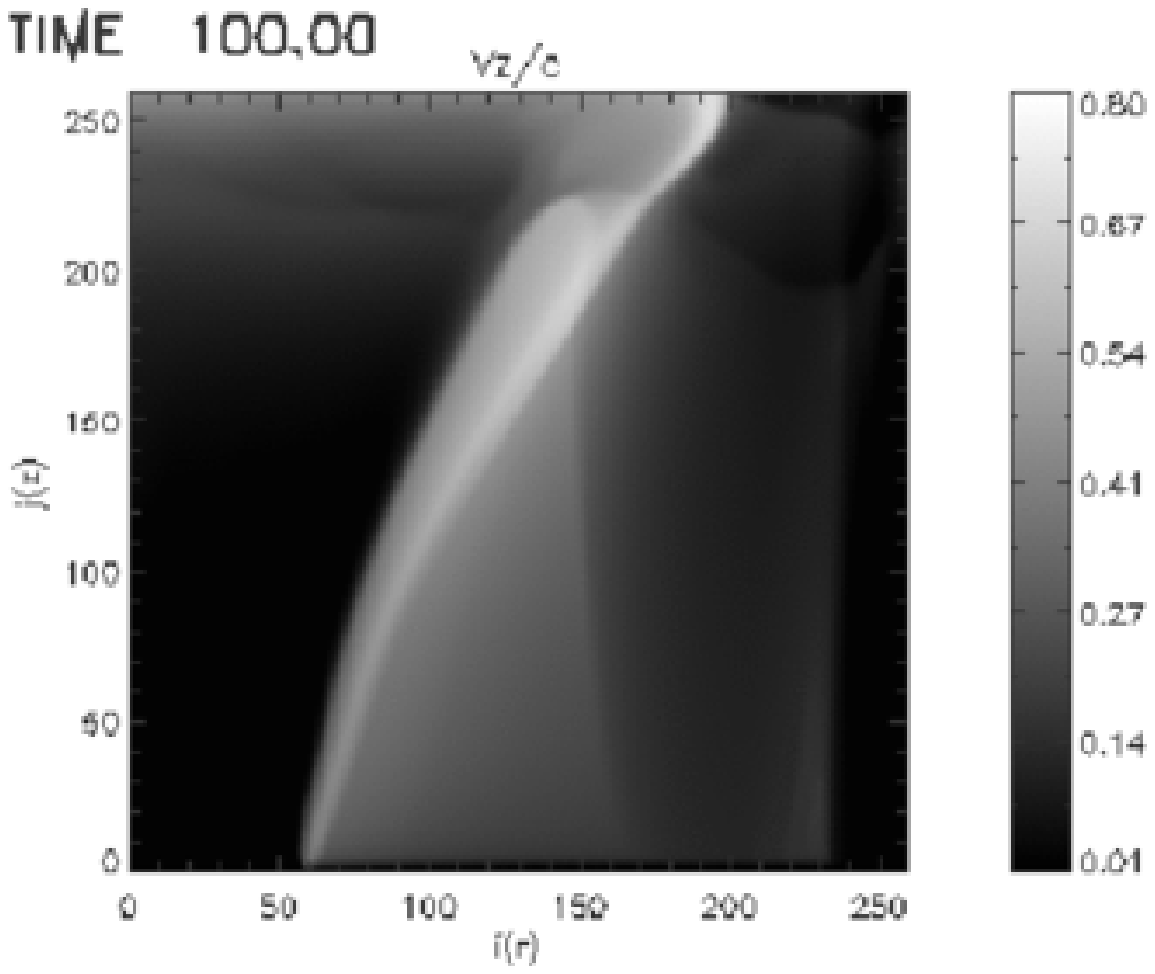,width=9truecm}
            \psfig{figure=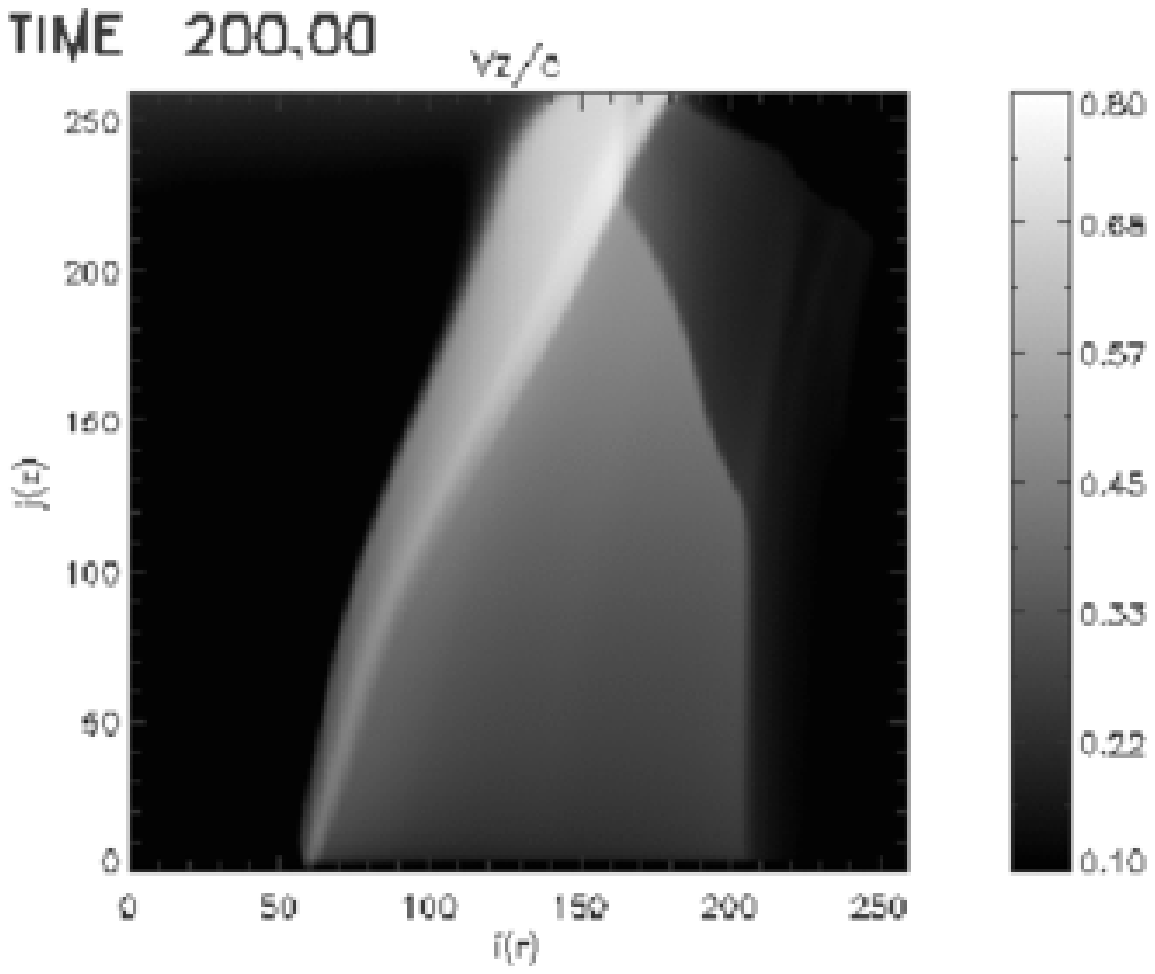,width=9truecm}}
\caption{Time evolution of a relativistic MHD jet simulation.
Shown is axial outflow velocity $v_z(r,z)$ normalized to the speed
of light $c$ at time steps 5, 10, 20, 50, 100, 200 (from {\it upper left}
to {\it lower right}).
Note the different color coding for different time steps.
Note that our plot is on a non-equidistant grid in order to resolve the
inner part of the jet. 
The grid points radii $r_i$ (and height $z_j$) for the cell (i,j) are 
located at
$r_{50} = 0.82, r_{60} = 1.00, r_{100} = 2.07, r_{150} = 5.21,
 r_{200} = 13.1,  r_{260} = 39.6$.
\label{fig:simul}}
\end{figure*}

\section{Thermal X-rays from relativistic jets}\label{sec:secXray}
Accretion disks around stellar mass black holes may reach temperatures 
up to $10^9$\,K close to their inner edge.
Assuming that this disk material becomes ejected into a disk wind 
and accelerated to a jet, 
each volume element of matter may radiate thermal Bremsstrahlung 
in the 1-10 keV X-ray band.
In case of optically thin emission, emission lines will be visible 
in the corresponding spectrum.
Relativistic motion of each volume element will boost and Doppler-shift 
its rest-frame spectrum. 
As the jet collimates, the line-of-sight angle of the moving elements
is changing, leading to a variation in the Doppler factor along the jet.

Figures \ref{fig:xray1} and \ref{fig:xray2} show example steps of
our derivation (see \cite{memo02} for details).
In the first step we calculate the MHD dynamics of a collimating jet
by solving the "hot" wind equation in Kerr metric from about 5 to 500
gravitational radii \cite{fend01a}. 
This provides temperature, density, velocity etc. along the flow.
In the second step we calculate the rest-frame spectrum of each of
the fluid elements along the jet, each having different temperature,
density (Fig.~\ref{fig:xray2}).
For thermal Bremsstrahlung the continuum is dominant for temperatures
$>10^9$\,K.
A number of lines appear for lower temperatures/densities
(e.g. lines of O, N (0.5-0.9 KeV), Ne, Fe, Mg (1-4 KeV) or
       FeXXV, FeXXVI emission (6.6-7.0 KeV).
In the last step we combine the local rest-frame spectra of all
fluid elements, 
each moving with a different velocity in respect to the line of
sight and, 
thus, affected by a different Doppler factor.
For our example of a micro-quasar we find a total X-ray luminosity 
in the rest frame of $4 \times 10^{31}$ ergs/s, 
respectively $6 \times 10^{32}$ ergs/s beamed and boosted and seen
from along the jet axis, or $2 \times 10^{33}$ ergs/s as seen from
20 degree inclination.
This is much less than the kinematic jet luminosity of 
$L_{\rm kin} \simeq 10^{39}$\,ergs/s (assuming mass loss
rates of $10^{-10}\,{\rm M_{\odot}}$/yr).

\section{Simulations of relativistic MHD jets}\label{sec:secMHDsim}
Finally we present preliminary results of numerical simulations 
considering the formation of relativistic MHD jets.
The overall goal is to test whether the paradigm of MHD self-collimation 
known for Newtonian jets \cite{ouye97, kras99, fend02b}
also holds in the relativistic case.

A Keplerian disk is taken as a fixed-in-time boundary condition with 
certain magnetic flux profile and mass loss rate from disk into corona.
This approach allows for long-term simulations of the evolving jet
which is the main advantage of {\em not} considering the disk evolution 
in the simulation. 
For simulations taking into account the disk structure we 
refer to the literature \cite{cass02, devi05, nish05, mcki07}.

Our ideal MHD simulations start from the initial state of a hydrostatic 
corona within a potential magnetic field (see \cite{fend08}).
We assume a polytropic gas with polytropic index $\gamma = 4/3$.
Turbulent magnetic diffusivity \cite{fend02b} or the variation of the 
magnetic flux and mass flow profiles across the jet \cite{fend06}
is not yet taken into account.
Mass is "injected" from the disk surface into the jet with low velocity 
of about 0.01 times the local Keplerian speed $v_{\rm K}(r)$.

We use the new MHD code PLUTO \cite{mign07} in the special relativity
option adding a Newtonian description of gravity.
The grid is non-equidistant with high resolution close to the jet axis
and the disk surface, spanning $40 \times 40$ inner disk radii $R_{\rm in}$.
The leading dynamical parameters for the simulations presented here are 
a plasma-$\beta = 0.1$, $v_{\rm K}(r=R_{\rm in}) = 0.4\,c$.

Figure \ref{fig:simul} shows our preliminary results. Shown is the axial component
of the velocity $v_z(r,z)/c$ for several time steps.
The low velocity material injected from the disk surface becomes accelerated to
about $0.8\,c$.
We observe two jet components - a collimated high speed beam emerging from the inner 
disk and an outer low velocity disk wind.
However, while the simulation reaches a quasi stationary state close to the jet axis,
the outer regions still dynamically evolve. 
This is natural as 100 rotations at the inner disk radius correspond to 0.5
rotations at the outer disk.

The whole outflow is still sub-Alfv\'enic with low toroidal magnetic field strength.
The degree of collimation as measured by the directed mass flux is about 5-10 degrees.
Since the flow is sub-Alfv\'enic we assume collimation by the surrounding low velocity
wind.

\balance

\end{document}